\documentclass[11pt]{amsart}
\usepackage{bbm,amsfonts,latexsym,amsmath,amscd,amssymb,fancyhdr}
\usepackage[all]{xy}
\usepackage{fullpage}

\theoremstyle{plain}
\newtheorem{theorem}{Theorem}
\numberwithin{theorem}{section}

\newtheorem{corollary}{Corollary}
\numberwithin{corollary}{section}

\numberwithin{definition}{section}

\numberwithin{lemma}{section}

\numberwithin{proposition}{section}

\newtheorem{remark}{Remark}
\numberwithin{remark}{section}

\numberwithin{example}{section}

\numberwithin{equation}{section}

\newcommand {\be}{\begin{equation}}
\newcommand {\ee}{\end{equation}}

\newcommand{\h}{\begin{eqnarray*}}
\newcommand{\e}{\end{eqnarray*}}

\newcommand{\CC}{\mathbf{C}}

\newcommand{\ZZ}{\mathbf{Z}}

\newcommand{\ii}{\sqrt{-1}}

\newcommand{\tr}{\mathrm{tr}}

\newcommand{\Tr}{\mathrm{Tr}}
\begin{document}

\title[Anomaly Cancellation and modularity. II: $E_8\times E_8$ case]{Anomaly cancellation and modularity II: the $E_8\times E_8$ case}
\author{Fei Han}
\address{Fei Han, Department of Mathematics, National University of Singapore,
 Block S17, 10 Lower Kent Ridge Road,
Singapore 119076 (mathanf@nus.edu.sg)}
\author{Kefeng Liu}
\address{Kefeng Liu, Department of Mathematics, University of California at Los Angeles,
Los Angeles, CA 90095, USA (liu@math.ucla.edu) and  Center of Mathematical Sciences, Zhejiang University, 310027, P.R. China}
\author{Weiping Zhang}
\address{Weiping Zhang, Chern Institute of Mathematics \& LPMC, Nankai
University, Tianjin 300071, P.R. China. (weiping@nankai.edu.cn)}
\maketitle

\begin{abstract} In this paper we  show that both of  the Green-Schwarz anomaly factorization formula for the gauge group $E_8\times E_8$ and the Ho\v{r}ava-Witten anomaly factorization formula for the gauge group $E_8$ can be  derived through  modular forms of weight 14.  This answers a question of   J. H. Schwarz. We also establish generalizations of these factorization formulas and obtain a new Ho\v{r}ava-Witten type factorization formula.
\end{abstract}

\section*{Introduction} In \cite{Liu1}, \cite{HLZ} and
 \cite{HLZ1}, it has been shown that both of the Alvarez-Gaum\'e-Witten miraculous anomaly cancellation formula \cite{AW} and the Green-Schwarz anomaly factorization formula \cite{GS} for the gauge group $SO(32)$  can be derived (and extended) through  a pair of modularly related modular forms, which are over the modular subgroup $\Gamma_0(2)$ and $\Gamma^0(2)$ respectively.  In answering a question of  J. H. Schwarz \cite{S1}, we deal with the remaining case   of gauge group $E_8\times E_8$ in this article.

Let $Z\to X\to B$ be a fiber bundle with fiber $Z$ being 10 dimensional.
Let $TZ$ be the vertical tangent bundle equipped with a metric
$g^{TZ}$ and an associated Levi-Civita connection $\nabla^{TZ}$
(cf. \cite[Proposition 10.2]{BGV}). Let  $R^{TZ}=(\nabla^{TZ})^2$
be the curvature of $\nabla^{TZ}$, which we also for  simplicity  denote  by $R$. Let $T_\CC Z$ be the
complexification of $TZ$ with the induced Hermitian connection
$\nabla^{T_\CC Z}$.

Let $(P_1, \vartheta_1), (P_2, \vartheta_2)$ be two principal $E_8$ bundles with connections over $X$.  Let $\rho$ be the adjoint representation of $E_8$. Let $W_i=P_i\times_\rho \CC^{248},\  i=1,\,2$ be the associated vector bundles, which are of rank $248$. We equip both $W_1$, $W_2$ with Hermitian metrics and Hermitian connections respectively. Let $F_i$ denote the curvature of the bundle $W_i$.  Let ``Tr" denote the trace in the adjoint representation. Then one has $\mathrm{Tr}F_i^{2n+1}=0$ (cf. the proof of Theorem 2.1 in this article), $\mathrm{Tr}F_i^{4}=\frac{1}{100}(\mathrm{Tr}F_i^{2})^2, \mathrm{Tr}F_i^{6}=\frac{1}{7200}(\mathrm{Tr}F_i^{2})^3$ (cf. \cite{Av}). It's easy to see that $c_2(W_i)=-\frac{1}{2}\mathrm{Tr}F_i^{2}.$ Simply denote $\Tr{F_1^n}+\Tr{F_2^n}$ by $\Tr{F^n}$.

The Green-Schwarz anomaly formula \cite{GS}  asserts that the following  factorization   for the 12 forms holds,\footnote{In what follows, we will write characteristic
forms without specifying the connections when there is no
confusion (cf. \cite{Z}).}
\be
\begin{split}
&I_{12}\\
=&\left\{\widehat{A}(TZ)\mathrm{ch}(W_1+W_2)+\widehat{A}(TZ)\mathrm{ch}(T_\CC Z)-2\widehat{A}(TZ)\right\}^{(12)}\\
=&\frac{-1}{64 \pi^6}\frac{1}{720}\left(-\frac{15}{8}\tr{R^2}\tr{R^4}-\frac{15}{32}(\tr R^2)^3+
\mathrm{Tr}F^6+\mathrm{Tr}F^2\left(\frac{1}{16} \tr{R^4}+\frac{5}{64}(\tr R^2)^2\right)-\frac{5}{8}\mathrm{Tr}F^4\tr{R^2} \right)\\
=&\frac{-1}{4 \pi^2}\frac{1}{2}\left(\tr{R^2}-\frac{1}{30}\Tr{F^2}\right)\cdot \frac{1}{16 \pi^4}\frac{1}{180}\left(\frac{1}{960}(\Tr F^2)^2-\frac{5}{16}\Tr F^4+\frac{1}{32}\tr R^2\Tr F^2-\frac{15}{16}\tr R^4-\frac{15}{64}(\tr R^2)^2 \right)\\
&\cdot \\
=&\left(p_1(TZ)+\frac{1}{30}(c_2(W_1)+c_2(W_2))\right)\cdot I_8.
\end{split}
\ee

In \cite{HW} and \cite{HW1},  Ho\v{r}ava and Witten  observed, on the other hand, that the following  anomaly factorization formula  holds for each  $i=1, 2$,
\be\begin{split}
\widehat{ I}^i_{12}=& \left\{\widehat{A}(TZ)\mathrm{ch}(W_i)+\frac{1}{2}\widehat{A}(TZ)\mathrm{ch}(T_\CC Z)-\widehat{A}(TZ)\right\}^{(12)}\\
=&\frac{-1}{64 \pi^6}\frac{1}{1440}\left(-\frac{15}{8}\tr{R^2}\tr{R^4}-\frac{15}{32}(\tr R^2)^3+
2\mathrm{Tr}F_i^6+\mathrm{Tr}F_i^2\left(\frac{1}{8} \tr{R^4}+\frac{5}{32}(\tr R^2)^2\right)-\frac{5}{4}\mathrm{Tr}F_i^4\tr{R^2} \right)\\
=&\frac{-1}{4 \pi^2}\frac{1}{4}\left(\tr{R^2}-\frac{1}{15}\Tr{F_i^2}\right)\cdot \widehat{I}^i_8\\
=&\left(\frac{1}{2}p_1(TZ)+\frac{1}{30}c_2(W_i)\right)\cdot \widehat{I}^i_8,
\end{split}
\ee
where $\widehat{I}^i_8$ can be written explicitly as
$$\widehat{I}^i_8=\frac{1}{16 \pi^4}\frac{1}{24}\left(-\frac{1}{4}\left(\frac{1}{2}\tr R^2-\frac{1}{30}\Tr F_i^2\right)^2-\frac{1}{8}\tr R^4+\frac{1}{32}(\tr R^2)^2 \right),$$
and therefore
$$I_{12}=\widehat{I}^1_{12}+\widehat{I}^2_{12}=\left(\frac{1}{2}p_1(TZ)+\frac{1}{30}c_2(W_1)\right)\cdot \widehat{I}^1_8+\left(\frac{1}{2}p_1(TZ)+\frac{1}{30}c_2(W_2)\right)\cdot \widehat{I}^2_8. $$

The purpose of this article is to show that the above   anomaly factorization formulas can also  be derived natually  from modularity as in the orthogonal group case dealt with in \cite{HLZ1}. This  provides a positive answer to a question of J. H. Schwarz mentioned at the beginning of the article.

To be more precse,
  we will construct in  Section 2 a modular form  $\mathcal{Q}(P_i, P_j, \tau)$ of weight 14 over $SL(2, \ZZ)$,  for any $i,\,j\in\{1,\,2\}$,   such that when $i=1, j=2$, the modularity of $\mathcal{Q}(P_1, P_2, \tau)$ gives the Green-Schwarz factorization formula (0.1), while when $i=j$, the modularity of $\mathcal{Q}(P_i, P_i, \tau)$ gives the Ho\v{r}ava-Witten factorization formula (0.2).  Actually what we construct is a more general modular form $\mathcal{Q}(P_i, P_j, \xi, \tau)$, which involves a complex line bundle (or equivalently a rank two real oriented bundle) and we are able to obtain generalizations of the Green-Schwarz formula and the Ho\v{r}ava-Witten formula by using the associated  modularity. Our construction of the modular form $\mathcal{Q}(P_i, P_j, \xi, \tau)$ involves the basic representation of the affine Kac-Moody algebra of $E_8$.

 Inspired by our modular method of deriving the Green-Schwarz and Ho\v{r}ava-Witten factorization formulas, we also construct a modular form $\mathcal{R}(P_i, \xi, \tau)$ of weight 10 over $SL(2, \ZZ)$, the modularity of which will give us a new factorization formula of  Ho\v{r}ava-Witten type. See Theorem 0.2 for details. It would be interesting to compare (0.8), (0.9) with the Ho\v{r}ava-Witten factorization (0.2) or (0.6).
 Actually another interesting question of J.H. Schwarz is to construct quantum field theories associated to the generalized anomaly factorization formulas in this paper and \cite{HLZ1}.

In the rest of this section, we will present our generalized Green-Schwarz and Ho\v{r}ava-Witten formula, as well as    the new formulas of Ho\v{r}ava-Witten type obtained from $\mathcal{R}(P_i, \xi, \tau)$. They will be   proved in Section 2 by using modularity after briefly reviewing some knowledge of the affine Kac-Moody algebra of $E_8$ in Section 1.

Let $\xi$ be a rank two real oriented Euclidean vector bundle over
$X$ carrying a Euclidean connection $\nabla^{\xi}$. Let $c=e(\xi,
\nabla^\xi)$ be the Euler form canonically associated to
$\nabla^\xi$ (cf. \cite[Section 3.4]{Z}).

\begin{theorem} The following identities hold, \be
\begin{split} &\left\{\widehat{A}(TZ)e^{\frac{c}{2}}\mathrm{ch}(W_1+W_2)+\widehat{A}(TZ)e^{\frac{c}{2}}\mathrm{ch}(T_\CC Z)-2\widehat{A}(TZ)e^{\frac{c}{2}}+\widehat{A}(TZ)e^{\frac{c}{2}}\mathrm{ch}(\widetilde{\xi_\CC}+3\widetilde{\xi_\CC}\otimes\widetilde{\xi_\CC}\right\}^{(12)}\\
=&\left(p_1(TZ)-3c^2+\frac{1}{30}(c_2(W_1)+c_2(W_2)) \right)\\
&\cdot \left\{-\frac{e^{\frac{1}{24}\left(p_1(TZ)-3c^2+\frac{1}{30}(c_2(W_1)+c_2(W_2)\right)} -1}{p_1(TZ)-3c^2+\frac{1}{30}(c_2(W_1)+c_2(W_2))}\widehat{A}(TZ)e^{\frac{c}{2}}\mathrm{ch}(\mathfrak{A})  +e^{\frac{1}{24}\left(p_1(TZ)-3c^2+\frac{1}{30}(c_2(W_1)+c_2(W_2)\right)}\widehat{A}(TZ)e^{\frac{c}{2}}\right\}^{(8)}
,\end{split}
\ee
where $\mathfrak{A}=W_1+W_2+T_\CC Z-2+\widetilde{\xi_\CC}+3\widetilde{\xi_\CC}\otimes\widetilde{\xi_\CC};$ \newline
and for each $i$,
\be
\begin{split} &\left\{\widehat{A}(TZ)e^{\frac{c}{2}}\mathrm{ch}(W_i)+\frac{1}{2}\widehat{A}(TZ)e^{\frac{c}{2}}\mathrm{ch}(T_\CC Z)-\widehat{A}(TZ)e^{\frac{c}{2}}+\frac{1}{2}\widehat{A}(TZ)e^{\frac{c}{2}}\mathrm{ch}(\widetilde{\xi_\CC}+3\widetilde{\xi_\CC}\otimes\widetilde{\xi_\CC})\right\}^{(12)}\\
=&\left(\frac{1}{2}p_1(TZ)-\frac{3}{2}c^2+\frac{1}{30}c_2(W_i) \right)\\
&\cdot \left\{-\frac{e^{\frac{1}{24}\left(p_1(TZ)-3c^2+\frac{1}{15}c_2(W_i)\right)} -1}{p_1(TZ)-3c^2+\frac{1}{15}c_2(W_i)}\widehat{A}(TZ)e^{\frac{c}{2}}\mathrm{ch}(\mathfrak{B}_i)  +e^{\frac{1}{24}\left(p_1(TZ)-3c^2+\frac{1}{15}c_2(W_i)\right)}\widehat{A}(TZ)e^{\frac{c}{2}}\right\}^{(8)}
,\end{split}
\ee
where $\mathfrak{B}_i=2W_i+T_\CC Z-2+\widetilde{\xi_\CC}+3\widetilde{\xi_\CC}\otimes\widetilde{\xi_\CC}. $
\end{theorem}

If $\xi$ is trivial, we obtain the Green-Schwarz formula (0.1) for $E_8\times E_8$ and the Ho\v{r}ava-Witten formula (0.2) for $E_8$ in the following corollary.
\begin{corollary} One has \be
\begin{split} &\left\{\widehat{A}(TZ)\mathrm{ch}(W_1+W_2)+\widehat{A}(TZ)\mathrm{ch}(T_\CC Z)-2\widehat{A}(TZ)\right\}^{(12)}\\
=&\left(p_1(TZ)+\frac{1}{30}(c_2(W_1)+c_2(W_2)) \right)\\
&\cdot \left\{-\frac{e^{\frac{1}{24}\left(p_1(TZ)+\frac{1}{30}(c_2(W_1)+c_2(W_2)\right)} -1}{p_1(TZ)+\frac{1}{30}(c_2(W_1)+c_2(W_2))}\widehat{A}(TZ)\mathrm{ch}(\mathfrak{C})  +e^{\frac{1}{24}\left(p_1(TZ)+\frac{1}{30}(c_2(W_1)+c_2(W_2)\right)}\widehat{A}(TZ)\right\}^{(8)}
,\end{split}
\ee
where $\mathfrak{C}=W_1+W_2+T_\CC Z-2;$ \newline
and for each $i$,
\be
\begin{split} &\left\{\widehat{A}(TZ)\mathrm{ch}(W_i)+\frac{1}{2}\widehat{A}(TZ)\mathrm{ch}(T_\CC Z)-\widehat{A}(TZ)\right\}^{(12)}\\
=&\left(\frac{1}{2}p_1(TZ)+\frac{1}{30}c_2(W_i) \right)\\
&\cdot \left\{-\frac{e^{\frac{1}{24}\left(p_1(TZ)+\frac{1}{15}c_2(W_i)\right)} -1}{p_1(TZ)+\frac{1}{15}c_2(W_i)}\widehat{A}(TZ)\mathrm{ch}(\mathfrak{D}_i)  +e^{\frac{1}{24}\left(p_1(TZ)+\frac{1}{15}c_2(W_i)\right)}\widehat{A}(TZ)\right\}^{(8)}
,\end{split}
\ee
where $\mathfrak{D}_i=2W_i+T_\CC Z-2. $
\end{corollary}
\begin{remark} It can be checked by direct computation that the second factors in the right hand sides of (0.5) and (0.6) are equal to $I_8$ and $\widehat{I}^i_8$ respectively. 
\end{remark}

We now state a new factorization formula, which is of  the Ho\v{r}ava-Witten type.

\begin{theorem} For each $i$, the following identity holds,
\be
\begin{split} &\left\{\widehat{A}(TZ)e^{\frac{c}{2}}\mathrm{ch}(W_i)+\widehat{A}(TZ)e^{\frac{c}{2}}\mathrm{ch}(T_\CC Z)+246\widehat{A}(TZ)e^{\frac{c}{2}}+\widehat{A}(TZ)e^{\frac{c}{2}}\mathrm{ch}(\widetilde{\xi_\CC}+3\widetilde{\xi_\CC}\otimes\widetilde{\xi_\CC})\right\}^{(12)}\\
=&\left(p_1(TZ)-3c^2+\frac{1}{30}c_2(W_i) \right)\\
&\cdot \left\{-\frac{e^{\frac{1}{24}\left(p_1(TZ)-3c^2+\frac{1}{30}c_2(W_i)\right)} -1}{p_1(TZ)-3c^2+\frac{1}{30}c_2(W_i)}\widehat{A}(TZ)e^{\frac{c}{2}}\mathrm{ch}(\mathfrak{E}_i)  +e^{\frac{1}{24}\left(p_1(TZ)-3c^2+\frac{1}{30}c_2(W_i)\right)}\widehat{A}(TZ)e^{\frac{c}{2}}\right\}^{(8)}
,\end{split}
\ee
where $\mathfrak{E}_i=W_i+T_\CC Z+246+\widetilde{\xi_\CC}+3\widetilde{\xi_\CC}\otimes\widetilde{\xi_\CC};$ \newline
if $\xi$ is trivial,  we have
\be
\begin{split} &\left\{\widehat{A}(TZ)\mathrm{ch}(W_i)+\widehat{A}(TZ)\mathrm{ch}(T_\CC Z)+246\widehat{A}(TZ)\right\}^{(12)}\\
=&\left(p_1(TZ)+\frac{1}{30}c_2(W_i) \right)\\
&\cdot \left\{-\frac{e^{\frac{1}{24}\left(p_1(TZ)+\frac{1}{30}c_2(W_i)\right)} -1}{p_1(TZ)+\frac{1}{30}c_2(W_i)}\widehat{A}(TZ)\mathrm{ch}(\mathfrak{F}_i)  +e^{\frac{1}{24}\left(p_1(TZ)+\frac{1}{30}c_2(W_i)\right)}\widehat{A}(TZ)\right\}^{(8)}
,\end{split}
\ee
where $\mathfrak{F}_i=W_i+T_\CC Z+246.$
\end{theorem}
\begin{remark} We can express (0.8) by direct computations as follows,
\be
\begin{split}
&\frac{-1}{64 \pi^6}\frac{1}{1440}\left(-\frac{15}{4}\tr{R^2}\tr{R^4}-\frac{15}{16}(\tr R^2)^3+
2\mathrm{Tr}F_i^6+\mathrm{Tr}F_i^2\left(\frac{1}{8} \tr{R^4}+\frac{5}{32}(\tr R^2)^2\right)-\frac{5}{4}\mathrm{Tr}F_i^4\tr{R^2} \right)\\
=&\frac{-1}{4 \pi^2}\frac{1}{2}\left(\tr{R^2}-\frac{1}{30}\Tr{F_i^2}\right)\cdot \frac{1}{16 \pi^4}\frac{1}{180}\left(-\frac{1}{480}(\Tr F_i^2)^2+\frac{1}{32}\tr R^2\Tr F_i^2-\frac{15}{16}\tr R^4-\frac{15}{64}(\tr R^2)^2 \right)\\
=&\left({p_1(TZ)+\frac{1}{30}c_2(W_i)}\right)\cdot \widehat{J}^i_8.
\end{split}
\ee
\end{remark}

\begin{remark} As in \cite{S1}, one may ask whether there is a phyics model corresponding to (0.8) and (0.9). 
\end{remark}

\section{The Basic Representation of Affine $E_8$}
In this section we   briefly review the basic representation theory for the affine $E_8$ by following \cite{K1} (see also \cite{K2}).

Let $\mathfrak{g}$ be the  Lie algebra of      $E_8$. Let $\langle, \rangle$ be the Killing form on $\mathfrak{g}$.
Let $\widetilde{\mathfrak{g}}$ be the affine Lie algebra corresponding to $\mathfrak{g}$ defined by
$$\widetilde{\mathfrak{g}}=\mathbf{C}[t, t^{-1}]\otimes \mathfrak{g}\oplus \CC c, $$ with bracket
$$[P(t)\otimes x+\lambda c, Q(t)\otimes y+\mu c]=P(t)Q(t)\otimes [x,y]+\langle x, y\rangle\,\mathrm{Res}_{t=0}\left(\frac{dP(t)}{dt}Q(t)\right)c.$$

Let $\widehat{\mathfrak{g}}$ be the affine Kac-Moody algebra obtained from $\widetilde{\mathfrak{g}}$ by adding a derivation $t\frac{d}{dt}$ which operates on $\mathbf{C}[t, t^{-1}]\otimes \mathfrak{g}$ in an obvious way and sends $c$ to $0$.

The basic representation $V(\Lambda_0)$ is the $\widehat{\mathfrak{g}}$-module defined by the property that there is a nonzero vector $v_0$ (highest weight vector) in $V(\Lambda_0)$ such that $cv_0=v_0, (\mathbf{C}[t]\oplus\CC t\frac{d}{dt})v_0=0$. Setting $V_k:=\{v\in V(\Lambda_0)| t\frac{d}{dt}=-kv\}$ gives a $\ZZ_+$-gradation by finite spaces. Since $[g,d]=0$, each $V_k$ is a representation of $\mathfrak{g}$. Moreover, $V_1$ is the adjoint representation of $E_8.$

Let $q=e^{2\pi\ii \tau}$. Fix a basis $\{z_i\}_{i=1}^8$ for the Cartan subalgebra. The character of the basic representation is given by
$$\mathrm{ch}(z_1, z_2,\cdots,z_8,\tau):=\sum_{k=0}^{\infty}(\mathrm{ch}V_k)(z_1, z_2,\cdots, z_8)q^k=\varphi(\tau)^{-r}\Theta_{\mathfrak{g}}(z_1, z_2,\cdots, z_8, \tau),$$
where $\varphi(\tau)=\prod_{n=1}^\infty (1-q^n)$ so that $\eta(\tau)=q^{1/24}\varphi(\tau)$ is the Dedekind $\eta$ function; $\Theta_{\mathfrak{g}}(z_1, z_2,\cdots, z_8, \tau)$ is the theta function defined on the root lattice $Q$ by
$$\Theta_{\mathfrak{g}}(z_1, z_2,\cdots, z_8, \tau)=\sum_{\gamma\in Q}q^{|\gamma|^2/2}e^{2\pi\ii \gamma(\overrightarrow{z})}.$$

It is proved in \cite{GL} (cf. \cite{Har}) that there is a basis for the $E_8$ root lattice such that
\be
\Theta_{\mathfrak{g}}(z_1, \cdots. z_8,\tau)
=\frac{1}{2}\left( \prod_{l=1}^8\theta(z_l,\tau)+\prod_{l=1}^8\theta_1(z_l,\tau)+\prod_{l=1}^8\theta_2(z_l,\tau)+\prod_{l=1}^8\theta_3(z_l,\tau)\right),
\ee
where $\theta$ and $\theta_i$ ($i=1,\ 2,\ 3$) are the Jacobi theta functions (cf. \cite{C} and \cite{HLZ}).

\section{Derivation of Green-Schwarz and Horava-Witten  type anomaly factorizations via modularity }
In this section, we will derive the Green-Schwarz and Ho\v{r}ava-Witten type factorization formulas in Theorems 0.1 and 0.2  via  modularity.

For the principal $E_8$ bundles $P_i, \ i=1,2,$ consider the associated bundles
$$\mathcal{V}_i=\sum_{k=0}^\infty \left(P_i\times_{\rho_k}V_k\right)q^k\in K(X)[[q]].$$ Since $\rho_1$ is the adjoint representation of $E_8$, we have $W_i=P_i\times_{\rho_1}V_1.$

Following \cite{CHZ}, set
\begin{equation*}
\Theta (T_{\mathbf{C}}Z,\xi_\mathbf{C}):= \left( \overset{
\infty }{\underset{m=1}{\otimes }}S_{q^{m}}(\widetilde{T_{\mathbf{C}}Z}
)\right) \otimes \left( \overset{\infty }{\underset{n=1}{\otimes }}\Lambda
_{q^{n}}(\widetilde{\xi_\CC})\right)
\otimes \left( \overset{\infty }{\underset{u=1}{\otimes }}\Lambda
_{-q^{u-1/2}}(\widetilde{\xi_\CC})\right) \otimes
\left( \overset{\infty }{\underset{v=1}{\otimes }}\Lambda _{q^{v-1/2}}(
\widetilde{\xi_\CC})\right)\in K(X)[[q]],
\end{equation*}
where $\xi_\CC$ is the complexification of $\xi$, and for a complex vector bundle $E$, $\widetilde{E}:=E-\CC^{{\rm rk}(E)}$.

Clearly, $\Theta(T_\CC Z, \xi_\CC)$ admits
a formal Fourier expansion in $q$ as \be \Theta(T_\CC Z,\xi_\CC)=\CC+B_1q+B_2q^2\cdots,\ee where
the $B_j$'s are elements in the semi-group formally generated by
complex vector bundles over $X$. Moreover, they carry canonically
induced connections denoted by $\nabla^{B_j}$. Let
$\nabla^{\Theta}$ be the induced connection with
$q$-coefficients on $\Theta$.

For $1\leq i, j \leq 2$,  set \be
\begin{split}
&\mathcal{Q}(P_i, P_j, \xi, \tau)\\
:=&\left\{e^{\frac{1}{24}E_2(\tau)\left(p_1(TZ)-3c^2+\frac{1}{30}(c_2(W_i)+c_2(W_j))\right)}\widehat{A}(TZ)\cosh\left(\frac{c}{2}\right)\mathrm{ch}\left(\Theta(T_\CC Z, \xi_\CC)\right)\varphi(\tau)^{16}\mathrm{ch}(\mathcal{V}_i)\mathrm{ch}(\mathcal{V}_j)\right\}^{(12)}.
\end{split}
\ee

\begin{theorem}$\mathcal{Q}(P_i, P_j, \xi, \tau)$ is a modular form of weight 14 over $SL(2, \ZZ)$.
\end{theorem}

\noindent {\it Proof}:
By the knowledge reviewed in Section 2, we see that there are formal two forms $y^i_l, 1\leq l \leq 8, i=1,2$ such that
\be \varphi(\tau)^{8}\mathrm{ch}(\mathcal{V}_i)=\frac{1}{2}\left(\prod_{l=1}^8\theta(y^i_l,\tau)+\prod_{l=1}^8\theta_1(y^i_l,\tau)+\prod_{l=1}^8\theta_2(y^i_l,\tau)+\prod_{l=1}^8\theta_3(y^i_l,\tau)\right).\ee

Since $\theta(z,\tau)$ is an odd function about $z$ and we only take forms of degrees not greater than12, one has
\be \varphi(\tau)^{8}\mathrm{ch}(\mathcal{V}_i)=\frac{1}{2}\left(\prod_{l=1}^8\theta_1(y^i_l,\tau)+\prod_{l=1}^8\theta_2(y^i_l,\tau)+\prod_{l=1}^8\theta_3(y^i_l,\tau)\right).\ee

Since $\theta_1(z,\tau), \theta_2(z,\tau)$ and $\theta_3(z,\tau)$ are all even functions about $z$, the right hand side of the above equality only contains even powers of $y^i_j$'s. Therefore $\mathrm{ch}(W_i)$ only consists of forms of degrees divisible by 4. So
\be  \mathrm{ch}(\mathcal{V}_i)=1+\mathrm{ch}(W_i)q+\cdots=1+(248-c_2(W_i)+\cdots)q+\cdots
.\ee

On the other hand,
\be \frac{1}{2}\left(\prod_{l=1}^8\theta_1(y^i_l,\tau)+\prod_{l=1}^8\theta_2(y^i_l,\tau)+\prod_{l=1}^8\theta_3(y^i_l,\tau)\right)=1+\left(240+30\sum_{l=1}^8(y^i_l)^2+\cdots\right)q+O(q^2).\ee

From (2.4), (2.5) and (2.6), we have
\be \sum_{l=1}^8\left(y^i_l\right)^2=-\frac{1}{30}c_2(W_i). \ee

Let $\{\pm 2\pi \ii x_l\}$ be the formal Chern roots for $(TZ_\CC, \nabla^{TZ_\CC})$. Let $c=2\pi \ii u$. One has
\be
\begin{split}
&\mathcal{Q}(P_i, P_j, \xi, \tau)\\
=&\left\{e^{\frac{1}{24}E_2(\tau)\left(p_1(TZ)-3c^2+\frac{1}{30}(c_2(W_i)+c_2(W_j)\right)}\widehat{A}(TZ)\cosh\left(\frac{c}{2}\right)\mathrm{ch}\left(\Theta(T_\CC Z, \xi_\CC)\right)\varphi(\tau)^{16}\mathrm{ch}(\mathcal{V}_i)\mathrm{ch}(\mathcal{V}_j)\right\}^{(12)}\\
=&\left\{e^{\frac{1}{24}E_2(\tau)\left(p_1(TZ)-3c^2+\frac{1}{30}(c_2(W_i)+c_2(W_j)\right)}\left(\prod_{l=1}^{5}\left(x_l\frac{\theta'(0,\tau)}{\theta(x_l,\tau)}\right)
\right)\frac{\theta_1(u,\tau)}{\theta_1(0,\tau)}\frac{\theta_2(u,\tau)}{\theta_2(0,\tau)}\frac{\theta_3(u,\tau)}{\theta_3(0,\tau)}\right.\\
& \left.\cdot \frac{1}{4}\left(\prod_{l=1}^8\theta_1(y^i_l,\tau)+\prod_{l=1}^8\theta_2(y^i_l,\tau)+\prod_{l=1}^8\theta_3(y^i_l,\tau)\right)\left(\prod_{l=1}^8\theta_1(y^j_l,\tau)+\prod_{l=1}^8\theta_2(y^j_l,\tau)+\prod_{l=1}^8\theta_3(y^j_l,\tau)\right)\right\}^{(12)}.
\end{split}
\ee

Then we can preform the transformation formulas for the theta functions and $E_2(\tau)$ (c.f. \cite{C} and \cite{HLZ}) to show that $\mathcal{Q}(P_i, P_j, \xi, \tau)$ is a modular form of weight 14 over $SL(2, \ZZ)$.
 Q.E.D.

$$ $$
\noindent {\it Proof of Theorem 0.1}:
Expanding the $q$-series, we have
\be
\begin{split}
&e^{\frac{1}{24}E_2(\tau)\left(p_1(TZ)-3c^2+\frac{1}{30}(c_2(W_i)+c_2(W_j)\right)}\widehat{A}(TZ)\cosh\left(\frac{c}{2}\right)\mathrm{ch}\left(\Theta(T_\CC Z, \xi_\CC)\right)\varphi(\tau)^{16}\mathrm{ch}(\mathcal{V}_i)\mathrm{ch}(\mathcal{V}_j)\\
=&\left(e^{\frac{1}{24}\left(p_1(TZ)-3c^2+\frac{1}{30}(c_2(W_i)+c_2(W_j))\right)}\right.\\
&\left.\ \ \ \ -e^{\frac{1}{24}\left(p_1(TZ)-3c^2+\frac{1}{30}(c_2(W_i)+c_2(W_j))\right)}\left(p_1(TZ)-3c^2+\frac{1}{30}(c_2(W_i)+c_2(W_j))\right)q+O(q^2)\right)\\
&\cdot \widehat{A}(TZ)\cosh\left(\frac{c}{2}\right)\mathrm{ch}(\CC+B_1q+O(q^2))(1-16q+O(q^2))(1+\mathrm{ch}(W_i)q+O(q^2))(1+\mathrm{ch}(W_j)q+O(q^2))\\
=&e^{\frac{1}{24}\left(p_1(TZ)-3c^2+\frac{1}{30}(c_2(W_i)+c_2(W_j))\right)}\widehat{A}(TZ)\cosh\left(\frac{c}{2}\right)\\
&+q\left(e^{\frac{1}{24}\left(p_1(TZ)-3c^2+\frac{1}{30}(c_2(W_i)+c_2(W_j))\right)}\widehat{A}(TZ)\cosh\left(\frac{c}{2}\right)
\mathrm{ch}(B_1-16+W_i+W_j)\right.\\
&\ \ \ \left.\ \ \ -e^{\frac{1}{24}\left(p_1(TZ)-3c^2+\frac{1}{30}(c_2(W_i)+c_2(W_j))\right)}\left(p_1(TZ)-3c^2+\frac{1}{30}(c_2(W_i)+c_2(W_j))\right)\widehat{A}(TZ)\cosh\left(\frac{c}{2}\right)\right)\\
&+O(q^2).
\end{split}
\ee

It is well known that modular forms over $SL(2, \ZZ)$ can be expressed as polynomials of the Eisenstein series $E_4(\tau)$, $E_6(\tau)$, where
\be E_4(\tau)=1+240q+2160q^2+6720q^3+\cdots,\ee
\be E_6(\tau)=1-504q-16632q^2-122976q^3+\cdots.\ee
Their weights are 4 and 6 respectively.

Since the weight of the modular form $\mathcal{Q}(P_i, P_j, \xi, \tau)$ is 14, it must be a multiple of
\be E_4(\tau)^2E_6(\tau)=1-24q+\cdots.\ee

So from (2.9) and (2.12), we have
\be
\begin{split}
&\left\{e^{\frac{1}{24}\left(p_1(TZ)-3c^2+\frac{1}{30}(c_2(W_i)+c_2(W_j))\right)}\widehat{A}(TZ)\cosh\left(\frac{c}{2}\right)
\mathrm{ch}(B_1-16+W_i+W_j)\right\}^{(12)}\\
&\ \ \ \ \ \ -\left\{e^{\frac{1}{24}\left(p_1(TZ)-3c^2+\frac{1}{30}(c_2(W_i)+c_2(W_j))\right)}\left(p_1(TZ)-3c^2+\frac{1}{30}(c_2(W_i)+c_2(W_j))\right)\widehat{A}(TZ)\cosh\left(\frac{c}{2}\right)\right\}^{(12)}\\
=&-24\left\{e^{\frac{1}{24}\left(p_1(TZ)-3c^2+\frac{1}{30}(c_2(W_i)+c_2(W_j))\right)}\widehat{A}(TZ)\cosh\left(\frac{c}{2}\right)\right\}^{(12)}.
\end{split}
\ee

Therefore
\be
\begin{split} &\left\{\widehat{A}(TZ)\cosh\left(\frac{c}{2}\right)\mathrm{ch}(W_i+W_j+B_1+8)\right\}^{(12)}\\
=&\left(p_1(TZ)-3c^2+\frac{1}{30}(c_2(W_i)+c_2(W_j)) \right)\\
&\cdot \left\{-\frac{e^{\frac{1}{24}\left(p_1(TZ)-3c^2+\frac{1}{30}(c_2(W_i)+c_2(W_j)\right)} -1}{p_1(TZ)-3c^2+\frac{1}{30}(c_2(W_i)+c_2(W_j))}\widehat{A}(TZ)\cosh\left(\frac{c}{2}\right)\mathrm{ch}(W_i+W_j+B_1+8) \right. \\
&\left.\ \ \ \ +e^{\frac{1}{24}\left(p_1(TZ)-3c^2+\frac{1}{30}(c_2(W_i)+c_2(W_j)\right)}\widehat{A}(TZ)\cosh\left(\frac{c}{2}\right)
\right\}^{(8)}
.\end{split}
\ee

To find $B_1$, we have
\be
\begin{split}
&\Theta (T_{\mathbf{C}}Z,\xi_\mathbf{C})\\
=& \left( \overset{
\infty }{\underset{m=1}{\otimes }}S_{q^{m}}(\widetilde{T_{\mathbf{C}}Z}
)\right) \otimes \left( \overset{\infty }{\underset{n=1}{\otimes }}\Lambda
_{q^{n}}(\widetilde{\xi_\CC})\right)
\otimes \left( \overset{\infty }{\underset{u=1}{\otimes }}\Lambda
_{-q^{u-1/2}}(\widetilde{\xi_\CC})\right) \otimes
\left( \overset{\infty }{\underset{v=1}{\otimes }}\Lambda _{q^{v-1/2}}(
\widetilde{\xi_\CC})\right)\\
=&(1+(T_\CC Z-10)q+O(q^2))\otimes(1+\widetilde{\xi_\CC}q+O(q^2))\\
&\otimes(1-\widetilde{\xi_\CC}q^{1/2}-2\widetilde{\xi_\CC}q+O(q^{3/2}))\otimes(1+\widetilde{\xi_\CC}q^{1/2}-2\widetilde{\xi_\CC}q+O(q^{3/2}))\\
=&1+(T_\CC Z-10+\widetilde{\xi_\CC}+3\widetilde{\xi_\CC}\otimes\widetilde{\xi_\CC})q+O(q^2).
\end{split}
\ee
So $B_1=T_\CC Z-10+\widetilde{\xi_\CC}+3\widetilde{\xi_\CC}\otimes\widetilde{\xi_\CC}.$

Plugging $B_1$ into (2.14), we have
\be
\begin{split} &\left\{\widehat{A}(TZ)\cosh\left(\frac{c}{2}\right)\mathrm{ch}(W_i+W_j+T_\CC Z-2+\widetilde{\xi_\CC}+3\widetilde{\xi_\CC}\otimes\widetilde{\xi_\CC} )\right\}^{(12)}\\
=&\left(p_1(TZ)-3c^2+\frac{1}{30}(c_2(W_i)+c_2(W_j)) \right)\\
&\cdot \left\{-\frac{e^{\frac{1}{24}\left(p_1(TZ)-3c^2+\frac{1}{30}(c_2(W_i)+c_2(W_j)\right)} -1}{p_1(TZ)-3c^2+\frac{1}{30}(c_2(W_i)+c_2(W_j))}\widehat{A}(TZ)\cosh\left(\frac{c}{2}\right)\mathrm{ch}(W_i+W_j+T_\CC Z-2+\widetilde{\xi_\CC}+3\widetilde{\xi_\CC}\otimes\widetilde{\xi_\CC}) \right. \\
&\left.\ \ \ \ +e^{\frac{1}{24}\left(p_1(TZ)-3c^2+\frac{1}{30}(c_2(W_i)+c_2(W_j)\right)}\widehat{A}(TZ)\cosh\left(\frac{c}{2}\right)
\right\}^{(8)}
.\end{split}
\ee

Since $\mathrm{ch}(W_i), \mathrm{ch}(W_j)$ only contribute degree $4l$ forms, we can replace $\cosh\left(\frac{c}{2}\right)$ by $e^{\frac{c}{2}}$. Then in (2.16), putting $i=1,\  j=2$ gives (0.4) and putting $i=j$ gives (0.5). Q.E.D.

$\ $

To prove theorem 0.2, for each $i$,  set
\be
\begin{split}
&\mathcal{R}(P_i, \xi, \tau)\\
:=&\left\{e^{\frac{1}{24}E_2(\tau)\left(p_1(TZ)-3c^2+\frac{1}{30}c_2(W_i)\right)}\widehat{A}(TZ)\cosh\left(\frac{c}{2}\right)\mathrm{ch}\left(\Theta(T_\CC Z, \xi_\CC)\right)\varphi(\tau)^{8}\mathrm{ch}(\mathcal{V}_i)\right\}^{(12)}.
\end{split}
\ee

\begin{theorem}$\mathcal{R}(P_i, \xi, \tau)$ is a modular form of weight 10 over $SL(2, \ZZ)$.
\end{theorem}
\noindent {\it Proof}: This can be similarly proved as Theorem 2.1 by seeing that
\be
\begin{split}
&\mathcal{R}(P_i, \xi, \tau)\\
=&\left\{e^{\frac{1}{24}E_2(\tau)\left(p_1(TZ)-3c^2+\frac{1}{30}c_2(W_i)\right)}\widehat{A}(TZ)\cosh\left(\frac{c}{2}\right)\mathrm{ch}\left(\Theta(T_\CC Z, \xi_\CC)\right)\varphi(\tau)^{8}\mathrm{ch}(\mathcal{V}_i)\right\}^{(12)}\\
=&\left\{e^{\frac{1}{24}E_2(\tau)\left(p_1(TZ)-3c^2+\frac{1}{30}c_2(W_i)\right)}\left(\prod_{l=1}^{5}\left(x_l\frac{\theta'(0,\tau)}{\theta(x_l,\tau)}\right)
\right)\frac{\theta_1(u,\tau)}{\theta_1(0,\tau)}\frac{\theta_2(u,\tau)}{\theta_2(0,\tau)}\frac{\theta_3(u,\tau)}{\theta_3(0,\tau)}\right.\\
& \left.\cdot \frac{1}{2}\left(\prod_{l=1}^8\theta_1(y^i_l,\tau)+\prod_{l=1}^8\theta_2(y^i_l,\tau)+\prod_{l=1}^8\theta_3(y^i_l,\tau)\right)\right\}^{(12)},
\end{split}\ee
and then apply the transformation laws of theta functions. Q.E.D.

$$ $$
\noindent {\it Proof of Theorem 0.2}: Similar as in the proof of Theorem 0.1,
expanding the $q$-series, we have
\be
\begin{split}
&e^{\frac{1}{24}E_2(\tau)\left(p_1(TZ)-3c^2+\frac{1}{30}c_2(W_i)\right)}\widehat{A}(TZ)\cosh\left(\frac{c}{2}\right)\mathrm{ch}\left(\Theta(T_\CC Z, \xi_\CC)\right)\varphi(\tau)^{8}\mathrm{ch}(\mathcal{V}_i)\\
=&\left(e^{\frac{1}{24}\left(p_1(TZ)-3c^2+\frac{1}{30}c_2(W_i)\right)}\right.\\
&\left.\ \ \ \ -e^{\frac{1}{24}\left(p_1(TZ)-3c^2+\frac{1}{30}c_2(W_i)\right)}\left(p_1(TZ)-3c^2+\frac{1}{30}c_2(W_i)\right)q+O(q^2)\right)\\
&\cdot \widehat{A}(TZ)\cosh\left(\frac{c}{2}\right)\mathrm{ch}(\CC+B_1q+O(q^2))(1-8q+O(q^2))(1+\mathrm{ch}(W_i)q+O(q^2))\\
=&e^{\frac{1}{24}\left(p_1(TZ)-3c^2+\frac{1}{30}c_2(W_i)\right)}\widehat{A}(TZ)\cosh\left(\frac{c}{2}\right)\\
&+q\left(e^{\frac{1}{24}\left(p_1(TZ)-3c^2+\frac{1}{30}c_2(W_i)\right)}\widehat{A}(TZ)\cosh\left(\frac{c}{2}\right)
\mathrm{ch}(B_1-8+W_i)\right.\\
&\ \ \ \left.\ \ \ -e^{\frac{1}{24}\left(p_1(TZ)-3c^2+\frac{1}{30}c_2(W_i)\right)}\left(p_1(TZ)-3c^2+\frac{1}{30}c_2(W_i)\right)\widehat{A}(TZ)\cosh\left(\frac{c}{2}\right)\right)\\
&+O(q^2).
\end{split}
\ee

However modular form of weight 10 must be a multiple of $E_4(\tau)E_6(\tau)=1-264q+\cdots$, so we have
\be
\begin{split}
&\left\{e^{\frac{1}{24}\left(p_1(TZ)-3c^2+\frac{1}{30}c_2(W_i)\right)}\widehat{A}(TZ)\cosh\left(\frac{c}{2}\right)
\mathrm{ch}(B_1-8+W_i)\right\}^{(12)}\\
&\ \ \ \ \ \ -\left\{e^{\frac{1}{24}\left(p_1(TZ)-3c^2+\frac{1}{30}c_2(W_i)\right)}\left(p_1(TZ)-3c^2+\frac{1}{30}c_2(W_i)\right)\widehat{A}(TZ)\cosh\left(\frac{c}{2}\right)\right\}^{(12)}\\
=&-264\left\{e^{\frac{1}{24}\left(p_1(TZ)-3c^2+\frac{1}{30}c_2(W_i)\right)}\widehat{A}(TZ)\cosh\left(\frac{c}{2}\right)\right\}^{(12)}.
\end{split}
\ee

Therefore
\be
\begin{split} &\left\{\widehat{A}(TZ)\cosh\left(\frac{c}{2}\right)\mathrm{ch}(W_i+B_1+256)\right\}^{(12)}\\
=&\left(p_1(TZ)-3c^2+\frac{1}{30}c_2(W_i) \right)\\
&\cdot \left\{-\frac{e^{\frac{1}{24}\left(p_1(TZ)-3c^2+\frac{1}{30}c_2(W_i)\right)} -1}{p_1(TZ)-3c^2+\frac{1}{30}c_2(W_i)}\widehat{A}(TZ)\cosh\left(\frac{c}{2}\right)\mathrm{ch}(W_i+B_1+256) \right. \\
&\left.\ \ \ \ +e^{\frac{1}{24}\left(p_1(TZ)-3c^2+\frac{1}{30}c_2(W_i)\right)}\widehat{A}(TZ)\cosh\left(\frac{c}{2}\right)
\right\}^{(8)}
.\end{split}
\ee

Plugging in $B_1$, we have
\be
\begin{split} &\left\{\widehat{A}(TZ)\cosh\left(\frac{c}{2}\right)\mathrm{ch}(W_i+T_\CC Z+246+\widetilde{\xi_\CC}+3\widetilde{\xi_\CC}\otimes\widetilde{\xi_\CC} )\right\}^{(12)}\\
=&\left(p_1(TZ)-3c^2+\frac{1}{30}c_2(W_i) \right)\\
&\cdot \left\{-\frac{e^{\frac{1}{24}\left(p_1(TZ)-3c^2+\frac{1}{30}c_2(W_i)\right)} -1}{p_1(TZ)-3c^2+\frac{1}{30}c_2(W_i)}\widehat{A}(TZ)\cosh\left(\frac{c}{2}\right)\mathrm{ch}(W_i+T_\CC Z+246+\widetilde{\xi_\CC}+3\widetilde{\xi_\CC}\otimes\widetilde{\xi_\CC}) \right. \\
&\left.\ \ \ \ +e^{\frac{1}{24}\left(p_1(TZ)-3c^2+\frac{1}{30}c_2(W_i)\right)}\widehat{A}(TZ)\cosh\left(\frac{c}{2}\right)
\right\}^{(8)}
.\end{split}
\ee

Since $\mathrm{ch}(W_i)$ only contribute degree $4l$ forms, we can replace $\cosh\left(\frac{c}{2}\right)$ by $e^{\frac{c}{2}}$, (2.22) gives us (0.7). Q.E.D.

$$ $$
\noindent {\bf Acknowledgements.} We are indebted to J. H.  Schwarz for asking us the question considered in this paper. The first author is partially supported by a start-up grant from National University of
Singapore. The second author is partially supported by NSF. The
third author is partially supported by  MOE and NNSFC.


\begin{thebibliography}{10}

\bibitem {Av} S.D. Avramis, Anomaly-free supergravities in six dimensions, {\it Arxiv:hep-th/0611133}.

\bibitem {AW} L. Alvarez-Gaum\'e and E. Witten, Gravitational anomalies, {\it Nucl. Physics}, B234  (1983), 269-330,



\bibitem{BGV}  N. Berline, E. Getzler and M. Vergne, {\it Heat
Kernels and Dirac Operators}. Springer, 2002.

\bibitem{C} K. Chandrasekharan, \textit{Elliptic Functions}.
Springer-Verlag, 1985.

\bibitem{CHZ} Q. Chen, F. Han and W. Zhang, Generalized Witten genus and vanishing theorems, \textit{J.   Diff. Geom.} 88
(2011), 1-39.

\bibitem{GL} T. Gannon and C. S. Lam, Lattices and $\theta$-function identities. II. Theta series. {\it J. Math. Phys.} 33  (1992), 871-887.


\bibitem{GS} M. B. Green  and  J. H. Schwarz, Anomaly cancellations in
supersymmetric d=10 gauge theory and superstring theory, {\it
Physics Letters} B149 (1984)   117-122.



\bibitem{HLZ} F. Han, K. Liu and W. Zhang,  Modular forms and generalized anomaly cancellation formulas, \textit{J. Geom. Phys.}, 62 (2012),   1038-1053.

\bibitem{HLZ1} F. Han, K. Liu and W. Zhang,  Anomaly cancellation and modularity. \textit{Arxiv: 1205.0718}


\bibitem{Har} C. Harris, The index bundle for a family of Dirac-Ramond operators, {\it Ph.D thesis}, University of Miami, 2012.




\bibitem{HW} P. Ho\v{r}ava and E. Witten, Heterotic and Type I string dynamics from eleven dimensions, {\it Nucl. Phys.} B460 (1996) 506.

\bibitem{HW1} P. Ho\v{r}ava and E. Witten, Eleven dimensional supergravity on a manifold with boundary, {\it Nucl. Phys.} B475 (1996) 94.

\bibitem{K1}  V. G. Kac,  An elucidation of : ``Infinite-dimensional algebras, Dedekind's
$\eta$-funtion, classical Moebius function and the very strange formula". $E_8{(1)}$ and the cube root of the modular function $j$. {\it Adv. in Math.}, 35(3): 264-273, 1980.

\bibitem{K2}  V. G. Kac, {\it Infinite-Dimensional Lie Algebras}, Cambridge University Press, Cambridge, 3rd edition, 1990.



\bibitem{Liu1} K. Liu, Modular invariance and characteristic numbers, {\it Comm. Math. Phys}, 174, 29-42 (1995).







\bibitem{S1} J. H. Schwarz, Private communications, 2012.


\bibitem {Z} W. Zhang, {\it Lectures on Chern-Weil Theory and
Witten Deformations.} Nankai Tracts in Mathematics Vol. 4, World
Scientific, Singapore, 2001.


\end{thebibliography}
\end{document}